\documentclass[modern]{rnaastex}

\newcommand{\ldl}{$\lambda/{\Delta}{\lambda}$}
\newcommand{\name}{WISE~J000100.45+065259.6}
\newcommand{\namesh}{WISE~J0001+0652}

\begin{document}

\title{Identification of WISE~J000100.45+065259.6 as an M8.5+T5 Spectral Binary Candidate}

%% Note that the corresponding author command and emails has to come
%% before everything else. Also place all the emails in the \email
%% command instead of using multiple \email calls.
\correspondingauthor{Adam J.\ Burgasser}
\email{aburgasser@ucsd.edu}

\author[0000-0002-6523-9536]{Adam J.\ Burgasser}
\affiliation{Center for Astrophysics and Space Science, University of California San Diego, La Jolla, CA 92093}

\author[0000-0002-9807-5435]{Christopher A.\ Theissen}
\affiliation{Center for Astrophysics and Space Science, University of California San Diego, La Jolla, CA 92093 }

\author[0000-0001-8170-7072]{Daniella C.\ Bardalez Gagliuffi}
\affiliation{American Museum of Natural History, Central Park West at 79th Street, New York, NY 10034 }

\author[0000-0001-8291-6490]{Everett Schlawin}
\affiliation{Steward Observatory, Tucson, AZ 85721}

%% Note that RNAAS manuscripts DO NOT have abstracts.

%% Start the main body of the article. If no sections in the 
%% research note leave the \section call blank to make the title.
\section{} 

Very low mass spectral binaries are systems composed of late-M/L dwarf primaries and T dwarf secondaries whose blended light spectra exhibit distinct peculiarities \citep{2010ApJ...710.1142B,2014ApJ...794..143B}. As spectral binaries are identified independent of separation, they provide a means of finding the smallest-separation low-mass stellar-brown dwarf pairs. 

\begin{figure}[h!]
\begin{center}
\includegraphics[trim={1cm 7cm 1cm 4.5cm},clip,scale=0.6,angle=0]{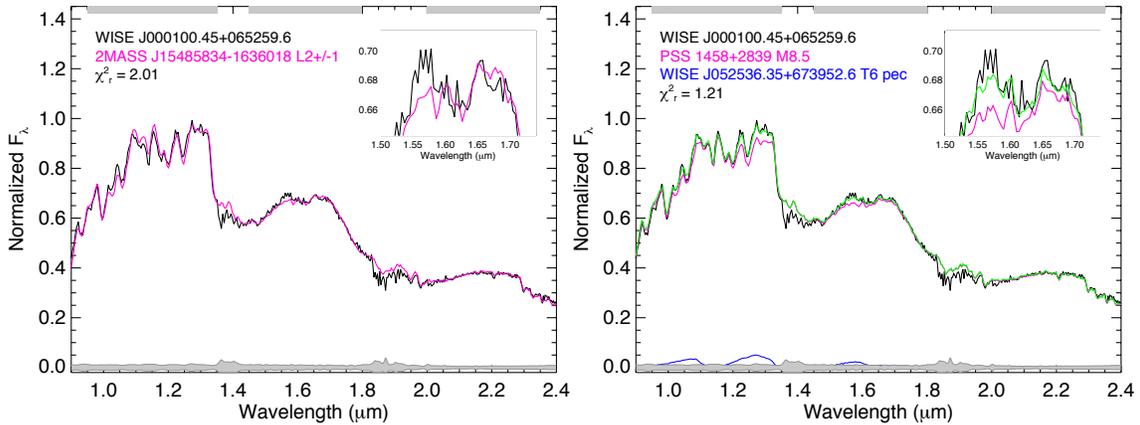}
\caption{Comparison of the IRTF/SpeX spectrum of {\namesh} (black lines) to best-fit single (left, in magnenta) and binary templates (right, primary in magneta, secondary in blue, combined in green). Uncertainty in the spectrum of {\namesh} is indicated by the grey band at bottom; fitting regions are indicated by the grey bands at top.
\label{fig:1}}
\end{center}
\end{figure}

As part of an ongoing program to spectroscopically confirm and classify bright late-M and L dwarfs,
 we observed {\name} (hereafter {\namesh}). This source was identified and photometrically classified as an L0 dwarf by \citet{2016A&A...589A..49S}, and independently found by \citet{2016AJ....151...41T,2017AJ....153...92T} as a high proper motion ($\mu$ = 114$\pm$12~mas/yr) late-type star at an estimated distance of 36~pc. To properly classify this source, we obtained low-resolution ({\ldl} $\approx$ 120), 0.8--2.45~$\micron$ spectra with the NASA 3m Infrared Telescope Facility SpeX spectrograph \citep{2003PASP..115..362R} on 2017 October 27 (UT) in partly cloudy conditions. The reduced spectrum (Figure~\ref{fig:1}; cf.\ \citealt{2003PASP..115..389V,2004PASP..116..362C}), is consistent with an early-type L dwarf, and comparison to spectral standards \citep{2010ApJS..190..100K}, indices \citep{2001AJ....121.1710R,2007ApJ...657..511A}, and spectral templates indicate a classification of M8--L2. However, specific peculiarities, such as excess flux at 1.25~$\micron$ and 1.57~$\micron$, are qualitatively similar to the blended light spectra of previously-discovered late-M plus T dwarf binaries (e.g., \citealt{2008ApJ...681..579B,2012ApJ...757..110B}), suggesting that {\namesh} is a very low mass spectral binary. 

We compared the spectrum of {\namesh} to 103,518 binary templates constructed from 729 M6-L2 primary and 142 L9-T6 secondary spectra, scaled to absolute fluxes using the \citet{2016ApJS..225...10F} field $M_J$/spectral type relation (cf.\ \citealt{2010ApJ...710.1142B}). 
%We avoided regions of strong telluric absorption at 1.35--1.45~$\micron$ and 1.8--2.0~$\micron$.
 Figure~\ref{fig:1} shows the best-fit binary template, which is a statistically significant improvement  (95\% confidence) over the best-fit single template based on the F-test statistic. A $\chi^2$-weighted average of the 296 best-fitting binary templates (P$_F$ $>$ 0.2) indicates component types of M8.5$\pm$0.5 and T5$\pm$1.2, with a magnitude difference of $\Delta{J}$ = 3.50$\pm$0.16.

Follow-up observations (high-resolution imaging, radial velocity and/or astrometric monitoring) are needed to confirm the binary nature of {\namesh}. As the majority of confirmed spectral binary candidates are very closely-separated systems ($\rho$ $\lesssim$ 3 AU; $P$ $\lesssim$ 15~yr; \citealt{2015AJ....150..163B}), {\namesh} may provide mass measurements within the decade. 

\clearpage

%% An example figure call using \includegraphics

\acknowledgments

\software{Astropy \citep{2013A&A...558A..33A},
	          SPLAT \citep{2017arXiv170700062B}}

%Acknowledge people, facilities, and software here but remember that this counts against your 1000 word limit.

%\bibliographystyle{aasjournal}
%\bibliography{biblibrary}

\end{document}